\documentclass[aps,prb,superscriptaddress]{revtex4}

\usepackage{graphicx}
\usepackage{amsfonts}
\usepackage{amssymb}
\usepackage[]{amsmath}
\usepackage[]{latexsym}
\usepackage[]{float}
\usepackage{bm}

\begin{document}


\title{Generalization of Chiral Symmetry for Tilted Dirac Cones}

\author{Tohru Kawarabayashi}
\affiliation{Department of Physics, Toho University,
Funabashi, 274-8510 Japan}

\author{Yasuhiro Hatsugai}
\affiliation{Institute of Physics, University of Tsukuba, Tsukuba, 305-8571 Japan}

\author{Takahiro Morimoto}
\affiliation{Department of Physics, University of Tokyo, Hongo, 
Tokyo 113-0033 Japan }

\author{Hideo Aoki}
\affiliation{Department of Physics, University of Tokyo, Hongo, 
Tokyo 113-0033 Japan }


\begin{abstract}
The notion of chiral symmetry for the conventional Dirac cone 
is generalized to include the tilted Dirac cones, where the generalized chiral operator turns out to be non-hermitian.
It is shown that the generalized chiral symmetry generically protects the zero modes ($n=0$ Landau level) of 
the Dirac cone even when tilted.  The present generalized symmetry is equivalent to 
the condition that the Dirac Hamiltonian is elliptic as a differential operator, 
which provides an explicit relevance to the index theorem. 
\end{abstract}


\maketitle

\section{Introduction}	

The chiral symmetry plays an important role 
in specifying some of the universality classes of the critical 
phenomena in disordered systems.\cite{LFSG,AZ,EM} 
Systems are called chiral-symmetric when there exists an operator $\Gamma$ 
that anti-commutes with the Hamiltonian, $\{ \Gamma , H\} =0$, with $\Gamma^\dagger = \Gamma$  and $\Gamma^2 =1$.
With this symmetry the energy eigenvalues appear  always in pairs $(E, -E)$, since if we have 
an eigenstate $\psi_E$ with an eigenvalue $E$, the state $\Gamma \psi_E$ is an 
eigenstate with an eigenvalue $-E$.  
The energy spectrum is therefore exactly particle-hole symmetric 
even when there exists a disorder as far as the disorder respects the chiral symmetry.  
In particular, the zero-energy state can be expressed as  an eigenstate of $\Gamma$.
  
For  a massless Dirac fermions in two dimensions, usually 
with vertical and isotropic Dirac cones as in graphene,\cite{Geim,Kim,Neto}  the effective Hamiltonian is expressed as  $H_0 = v_F (\sigma_x \pi_x + \sigma_y \pi_y)$, 
where $\sigma_{x(y)}$ is the Pauli matrix and $\bm{\pi} = \bm{p} + e\bm{A}$ denotes the dynamical momentum with 
the vector potential $\bm{A}$ and the electron charge $-e$.  The fermi velocity is denoted by $v_F$.
For such a vertical Dirac cone, we have obviously $\{ H_0 ,\sigma_z\} =0$ and thus the system is chiral-symmetric with $\Gamma = \sigma_z$. The zero-energy Landau level (zero modes), 
which is essential to the anomalous quantum Hall effect 
for massless Dirac fermions in a magnetic field, then becomes an eigenstate of $\sigma_z$. 
The most remarkable property with these zero modes is their robustness against disorder in gauge 
degrees of freedom. The zero energy ($n=0$) Landau level does not acquire any width due to such a 
disorder, while other Landau levels are broadened as usual, and this gives rise to   an unconventional 
criticality for the quantum Hall transition at the $n=0$ Landau level.\cite{OGM}
This robustness of zero modes for a vertical Dirac cone 
has been discussed in terms of the index theorem\cite{Nakahara,KN} or based on the explicit form of wave functions due to 
Aharonov and Casher.\cite{AC,Kail} It has been also demonstrated numerically that the chiral symmetry 
is also crucial to this robustness of zero modes of vertical Dirac cones.\cite{KHA}

Massless Dirac fermions in two dimensions appear not only in graphene but also in certain organic metals, where 
we encounter pairs of tilted Dirac cones.\cite{KKS,TSTNK,MHT,MT,GFMP} The existence of zero modes and 
the Landau level structure have been established also for tilted  Dirac cones. However, the effect of disorder, in particular the 
robustness of zero modes and the role of symmetry, has not been explored until recently.\cite{KHMA}
In the present paper, with an explicit form of the eigenstates of the generalized chiral operator, 
we demonstrate how the chiral symmetry, which is broken for tilted cones, can be generalized to include
tilted Dirac cones and clarify its relevance to the robustness of the zero modes of generic massless Dirac fermions.  Relationships between 
the generalized chiral symmetry and the applicability of the index theorem  is also elaborated.

\section{Generalization of Chiral Symmetry}

To illustrate how the chiral symmetry is generalized to tilted Dirac cones, 
let us consider a general form of the Hamiltonian,\cite{KHMA}
$$
 H(\eta) = -\eta v_F \sigma_0 \pi_x + v_F(\sigma_x \pi_x  +\sigma_y\pi_y),
$$
for a two-dimensional massless Dirac fermion in a magnetic field, where  the isotropic Dirac cone 
is tilted in the $x$ direction for $\eta \neq 0$ 
with $\sigma_0$ being the unit matrix.  
The dynamical momentum $\bm{\pi}$ satisfies the commutation relation 
$[\pi_x,\pi_y] = -ie\hbar B$ with $\bm{B} = {\rm rot} \bm{A}$. The parameter $\eta$ determines the tilting of the Dirac cone.
Note that for $\bm{A}=0$ an equienergy contour of the Dirac cone 
is elliptic as long as $|\eta|<1$ (while hyperbolic for $|\eta|>1$;  see 
Fig.\ref{f1}(a)). 
The first term in the Hamiltonian destroys the chiral symmetry as $\sigma_z H(\eta) \sigma_z = -H(-\eta) \neq -H(\eta)$. 
We can, however,  define a generalized chiral operator $\gamma$ as
$$
 \gamma = \lambda_\eta^{-1}(\sigma_z -i \eta  \sigma_y) , \quad \lambda_\eta = \sqrt{1-\eta^2},
$$
which arises naturally in the general framework of the eigenvalue problem for tilted Dirac cones.\cite{HKA}
Although the generalized chiral operator $\gamma$ is not hermitian ($\gamma^\dagger \neq \gamma$), one can readily verify that 
$\gamma^2 =1$ with eigenvalues $\pm 1$. The corresponding right-eigenvectors $\gamma |\pm \rangle = \pm |\pm \rangle$ are 
given explicitly as  
$$
 | + \rangle = \frac{1}{\sqrt{2(1+\lambda_\eta)}}\left(\begin{array}{c} 
 1+ \lambda_\eta \\
 \eta 
 \end{array}\right) , \quad
 | - \rangle = \frac{1}{\sqrt{2(1+\lambda_\eta)}}\left(\begin{array}{c} 
 \eta \\
1+\lambda_\eta 
 \end{array}\right).
$$
In the limit of the vertical cone ($\eta \to 0$), the generalized operator $\gamma$
reduces to the conventional chiral operator $ \sigma_z$.
With the generalized chiral operator, we find that for $|\eta| <1$,
$$
 \gamma^\dagger H(\eta) \gamma = -H(\eta),
$$
which we call the {\it generalized chiral symmetry}.\cite{KHMA} 
The generalized symmetry guarantees 
the identity $ \langle + | H(\eta) | + \rangle =  \langle - | H(\eta) | - \rangle =0$, and plays an essential role for the robustness of zero modes as shown in the next section.
Note that this symmetry holds irrespective of the details of the vector potential $\bm{A}$.
Disorder in gauge degrees of freedom (such as random magnetic fields) 
respects this symmetry. 

\section{Robustness of Zero Modes}

If the above generalized chiral symmetry is preserved,
the Schr\"{o}dinger equation $H(\eta)\psi  = E\psi$  for the wave function $\psi = |+\rangle \psi^+ + |-\rangle \psi^- $ becomes 
$$
 \left(\begin{array}{cc}
 0 & \lambda_\eta \pi_x -i \pi_y \\
\lambda_\eta \pi_x+i\pi_y & 0
 \end{array}\right)
 \left(\begin{array}{c}
 \psi^+ \\
 \psi^-
 \end{array}\right) = 
E \left(\begin{array}{cc}
 \ 1/\lambda_\eta\  &\ \eta/\lambda_\eta\  \\
 \ \eta/\lambda_\eta\  & \ 1/\lambda_\eta\ 
 \end{array}\right)
 \left(\begin{array}{c}
 \psi^+ \\
 \psi^-
 \end{array}\right) .
$$
The zero ($E=0$) modes are then given by the wave functions satisfying\cite{KHMA}
\begin{equation}
 (\lambda_\eta \pi_x - i \pi_y)\psi^- =0 , \  \psi^+ =0 
 \label{eq1}
\end{equation}
or
\begin{equation}
 (\lambda_\eta \pi_x + i \pi_y)\psi^+ =0 , \  \psi^- =0 .
\label{eq2}
\end{equation}
The zero modes are thus the eigenstates of the generalized chiral operator $\gamma$ and have either ``$-$" chirality with $\gamma \psi =- \psi$ (Eq.(\ref{eq1})) or  ``$+$" chirality with $\gamma \psi = \psi$ (Eq.(\ref{eq2})). It is to be recalled that 
for a vertical Dirac cone the zero modes are also the eigenstates of the chiral operator $\Gamma=\sigma_z$. 
These equations  for the zero modes hold even in the case where the gauge field is disordered.

Following Aharonov and Casher,\cite{AC} we adopt the ``Coulomb gauge" $\lambda_\eta \partial_x A_x+\lambda_\eta^{-1}\partial_y A_y=0$ by 
assuming $\bm{A} = (-\lambda_\eta^{-1}\partial_y \varphi, \lambda_\eta \partial_x \varphi)$. Then Eq.(\ref{eq1}) for $\psi^-$ and Eq.(\ref{eq2}) for $\psi^+$ are 
reduced to 
$$
 [D_\pm \mp (e/\hbar)(D_\pm\varphi)]\psi^\pm =0,
$$
where $D_\pm \equiv (\partial_X \pm i\partial_Y)$ with $\bm{R}=(X,Y)=(x/\sqrt{\lambda_\eta}, y\sqrt{\lambda_\eta} )$.
The solutios are then given by $\psi^\pm = \exp(\pm e\varphi/\hbar)f(Z_\pm)$ with  a polynomial $f(Z_\pm)$ in $Z_{\pm} \equiv X \pm iY$. 
Since $(\partial_X^2 + \partial_Y^2) \varphi =B$, we have $\varphi (\bm{R}) = \int d\bm{R}' G(\bm{R}-\bm{R}')B(\bm{R}')$ where $G(\bm{R}) = 
(1/2\pi)\log (R)$ with $R=\sqrt{X^2+Y^2}$, which leads to the asymptotic form  $\varphi \to (\Phi/2\pi)\log R$ in the 
limit as $R \to \infty$ with $\Phi$  being the total magnetic flux in the system. 
We then have to chose the chirality for the wave function to be normalizable. For instance, when the total 
magnetic flux $\Phi$ is positive, the zero mode has to have ``$-$" chirality to be normalizable.
The number of square-normalizable wave function  then
becomes $\Phi/(h/e)$,\cite{AC}  that exactly coincides with the degeneracy of the Landau level.  This implies the zero-modes exhaust the Landau level of a tilted Dirac cone so that its density of states is a {\it generalized chiral symmetry protected delta-function} in the presence of disorder.  
In this sense, the generalized chiral symmetry protects the zero mode
of a generic  massless Dirac fermions in two dimensions.

\section{Generalized Chiral Symmetry and Index Theorem}

The above reasoning gives an explicit relationship between the index theorem and the generalized 
chiral symmetry. The index theorem, which gives the least  number of zero modes,  holds for an elliptic operator.\cite{Nakahara}
In the present case, the Hamiltonian $H(\eta)$ is elliptic as a differential operator if the matrix,
$$
 \Xi(\xi_x,\xi_y)=\left(\begin{array}{cc}
 -\eta \xi_x &\xi_x -i \xi_y \\
 \xi_x +i\xi_y & -\eta \xi_x
 \end{array}\right),
$$
is invertible for any $(\xi_x,\xi_y) \in {\rm \bf R^2}-{(0,0)}$. The condition for the ellipticity thus becomes $\det \Xi = -(1-\eta^2)\xi_x^2 - \xi_y^2  \neq 0$. 
This determinant is always negative ($\det \Xi <0$) and never becomes 
zero for $(\xi_x,\xi_y) \in {\rm \bf R^2}-{(0,0)}$, as long as $|\eta|<1$.
The condition $|\eta|<1$ is therefore exactly  the condition for the ellipticity of 
the Hamiltonian $H(\eta)$ as well as that for the existence of the generalized chiral symmetry.
In this sense, {\it the generalized chiral symmetry is generically 
equivalent to the ellipticity of the 
Dirac-cone Hamiltonian} with the same parameter space for their 
validity. 
Geometrically, the condition $|\eta|<1$ means that the Dirac cone is not tilted too much 
so that the cross section with a constant energy plane remains to be elliptic.

\section{Numerical Demonstration} 

\begin{figure}[h]
\includegraphics[width=13cm]{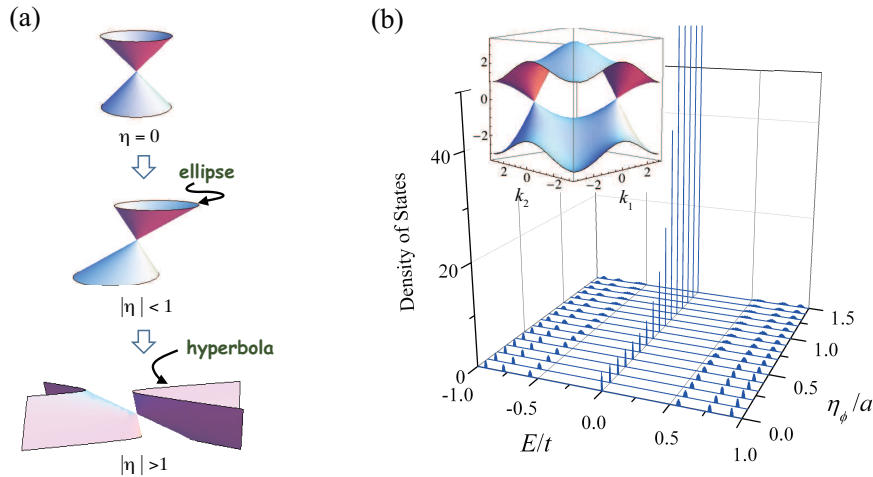}
\vspace*{8pt}
\caption{(a) Schematic figures of the Dirac cones with $\eta=0$, $|\eta|<1$ and $|\eta|>1$. 
(b) Density of states for the lattice model in a magnetic field with various values of the correlation length $\eta_\phi$ of 
the random component of the magnetic flux. Here, the parameters are assumed to be $t'/t=0.25$, $\phi=0.01(h/e)$, and 
$\sqrt{ \langle \delta\phi^2 \rangle}= 0.0029(h/e)$. The system-size is 20$a$ by 20$a$ and the average over $10^4$ samples is made.
Inset: Energy dispersion $E(\bm{k})/t$ in the absence of a magnetic field, where  $k_1= \bm{k}\cdot (\bm{e}_x + \bm{e}_y)$ and
$k_2= \bm{k}\cdot (\bm{e}_x - \bm{e}_y)$. 
\label{f1}}
\end{figure}

We have also preformed numerical calculations based on a lattice model  having 
a pair of tilted Dirac cones at $E=0$ as shown in Fig.\ref{f1}(b)(Inset). The model is defined on the two-dimensional square lattice
with a Hamiltonian having the nearest-neighbor $(t)$ and the next-nearest neighbor $(t')$ transfer integrals as \cite{KHMA,piflux}
$$
 H = \sum_{\bm{r}}-tc_{\bm{r}+\bm{e}_x}^\dagger c_{\bm{r}} +(-1)^{x+y}tc_{\bm{r}+\bm{e}_y}^\dagger c_{\bm{r}} +t'(
 c_{\bm{r}+\bm{e}_x+\bm{e}_y}^\dagger c_{\bm{r}} +c_{\bm{r}+\bm{e}_x-\bm{e}_y}^\dagger c_{\bm{r}})+ {\rm H.c.},
$$
where $\bm{r}=(x,y)$ denotes the lattice point in units of the nearest-neighbor distance $a$, and $\bm{e}_x(\bm{e}_y)$ the unit vector 
in the $x(y)$ direction. The magnetic field is taken into account by the Peierls phases $\theta(\bm{r})$ as $t(t') \to t(t')\exp(-2\pi i\theta(\bm{r}))$, so that
the their sum along the closed loop is equal to the enclosed magnetic flux  in units of the flux quantum $h/e$. 
To see how the Landau levels are broadened by disorder in the present model, we assume that   
the magnetic flux $\phi(\bm{r})$ enclosed by the square located at $\bm{r}$ has a random component $\delta \phi (\bm{r})$
in addition to the uniform component $\phi$. The random components have  a gaussian distribution and  are correlated in space 
as $\langle \delta\phi(\bm{r}_1) \delta\phi(\bm{r}_2) \rangle =  \langle \delta\phi^2 \rangle \exp(-|\bm{r}_1-\bm{r}_2|^2/4\eta_\phi^2)$.
The disorder in magnetic fluxes should appear, for large $\eta_\phi$, as the disorder in gauge 
degrees of freedom, and therefore should respect the generalized chiral symmetry of  the effective Hamiltonian at the Dirac points. 
As shown in Fig.\ref{f1}(b), we actually see an anomalously sharp $n=0$ Landau level when the correlation length $\eta_\phi$ of disorder becomes larger than the nearest-neighbor distance $a$,
while other Landau levels ($n=\pm1,\pm2,\pm3$) are broadened as usual. 
Together with the result\cite{KHMA} for a different inclination of the cone, this anomaly at the $n=0$ 
Landau level is likely to exist universally, irrespective of the magnitude of the tilting angle of the cone. 
This anomalously sharp $n=0$ Landau level  suggests that 
the energy levels are exactly degenerated at $E=0$. The present numerical results are thus 
consistent with the robustness of zero modes of a single tilted Dirac cone 
protected by the generalized chiral symmetry.

\section{Conclusions}

We have shown explicitly that the notion of the chiral symmetry can be generalized to generic tilted Dirac cones, where 
the generalized  chiral operator has to be non-hermitian. It has been also demonstrated analytically that
the generalized chiral symmetry indeed protects the zero modes of the system by extending the 
argument by Aharonov and Casher.\cite{AC} The resulting anomalously sharp $n=0$  Landau level 
has been also confirmed numerically  based on the lattice model. The existence of the generalized chiral symmetry 
coincides with the ellipticity of the Hamiltonian as a differential operator, which is nothing but the geometrical 
condition that the Dirac cone is not tilted too much so that the cross section with a constant energy plane is an ellipse.

\section*{Acknowledgments}

The authors wish to thank Yoshiyuki Ono and Tomi Ohtsuki for useful discussions.
This work was partly supported by Grants-in-Aid for Scientific Research, Nos. 22540336 and 23340112 from JSPS.


\vfill
\end{document}